\documentclass[twoside,twocolumn,english,prl, showpacs]{revtex4}

\usepackage[latin1]{inputenc}
\usepackage{graphicx}
\usepackage{amssymb}

\makeatletter

\usepackage{babel}
\makeatother
\begin{document}

\title{Fermionic Mach-Zehnder interferometer subject to a quantum bath}

\author{Florian Marquardt}

\affiliation{Departments of Physics and Applied Physics, Yale University, PO Box
208284, New Haven, CT 06511, USA}

\date{12.10.2004}

\email{Florian.Marquardt@yale.edu}

\begin{abstract}
We study fermions in a Mach-Zehnder interferometer, subject to a quantum-mechanical
environment leading to inelastic scattering, decoherence, renormalization
effects, and time-dependent conductance fluctuations. Both the loss
of interference contrast as well as the shot noise are calculated,
using equations of motion and leading order perturbation theory. The
full dependence of the shot-noise correction on setup parameters,
voltage, temperature and the bath spectrum is presented. We find an
interesting contribution due to correlations between the fluctuating
renormalized phase shift and the output current, discuss the limiting
behaviours at low and high voltages, and compare with simpler models
of dephasing.
\end{abstract}

\pacs{73.23.-b, 72.70.+m, 03.65.Yz}

\maketitle
\emph{Introduction} - Quantum interference effects and their destruction
by scattering play a prominent role in mesoscopic physics. In contrast
to the usual Aharonov-Bohm ring setups, the recently introduced Mach-Zehnder
interferometer for electrons\cite{HeiblumEtAl} offers an exciting
possibility to study an ideal two-way interference geometry, with
chiral single-channel transport and in the absence of backscattering.
The loss of visibility with increasing bias voltage or temperature
has been observed, and the idea of using shot noise measurements to
learn more about potential dephasing mechanisms has been introduced.

On the theoretical side, the loss of interference contrast in the
current had been studied for the Mach-Zehnder setup\cite{Seelig}
prior to this experiment. More recently the influence of dephasing
on shot noise has been analyzed\cite{unserPRL}, revealing important
differences between phenomenological and microscopic approaches. However,
both of these works deal with a classical noise field acting on the
electrons. Thus, issues such as the increase of the dephasing rate
with rising bias voltage could not be studied, as this effect is due
to lifting the restrictions of Pauli blocking on the scattering of
particles, which do not apply to classical noise. 

\begin{figure}
\includegraphics[%
  width=1.0\columnwidth]{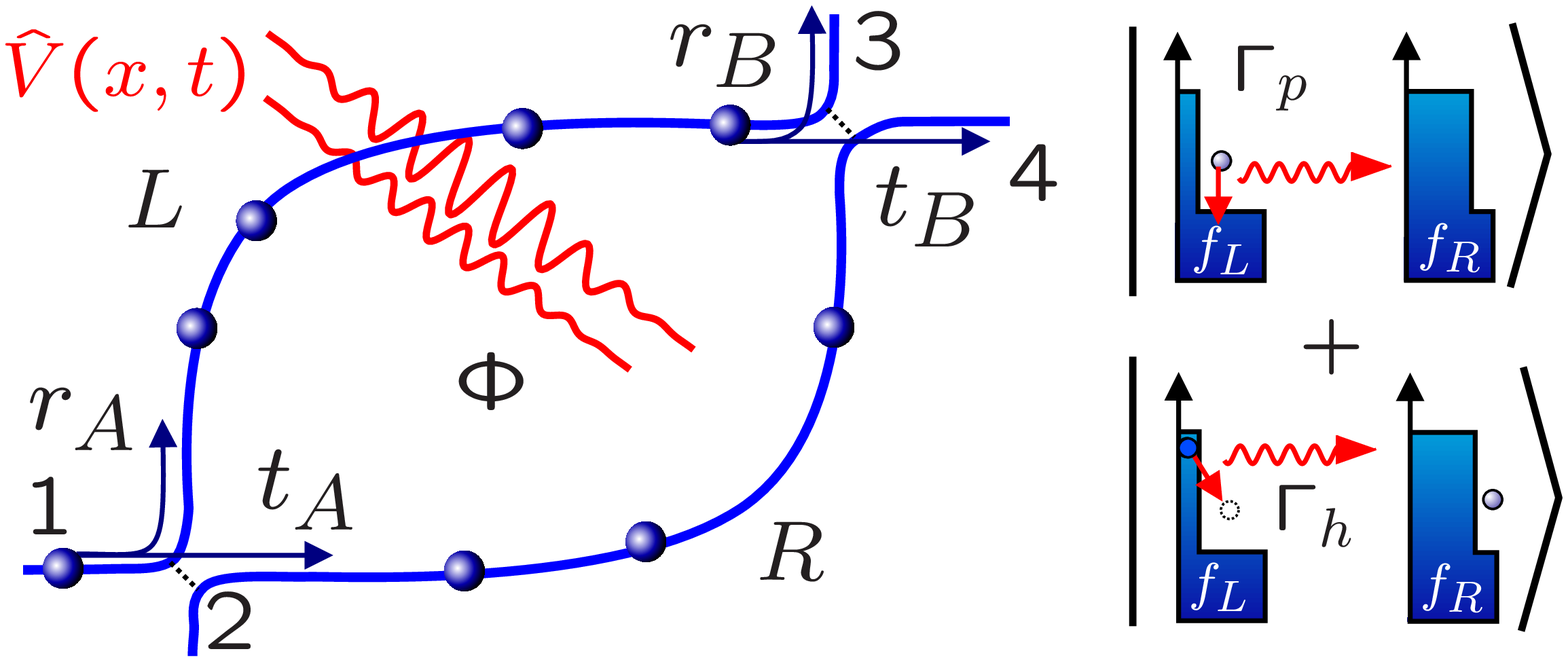}

\caption{\label{cap:Schematic-of-the}Left: Schematic of the Mach-Zehnder
setup. Right: Combination of particle- and hole-scattering processes
contributing to the dephasing rate.}
\end{figure}

In this work, we study the influence of any true quantum bath (phonons,
Nyquist noise, etc.) on a fermionic Mach-Zehnder interferometer (Fig.
\ref{cap:Schematic-of-the}). Besides its experimental relevance \cite{HeiblumEtAl},
this setup represents an ideal model problem in which many of the
features of a quantum bath acting on a fermion system can be analyzed
more easily and/or thoroughly than in more complicated situations
such as weak localization\cite{WLdephasingClassical,WLdephasingDiags}.
We fully account for Pauli blocking in a nonequilibrium transport
situation (i.e. arbitrary bias) and derive both the dephasing rate,
as well as the effects on the current noise. We employ a physically
transparent equations of motion approach that is analogous to the
case of classical noise, but keeps the Pauli principle via the backaction
of the bath onto the system. The evaluation will be performed perturbatively,
to leading order in the system-bath interaction.

\emph{The model} - We consider a model of spinpolarized fermions,
moving chirally and without backscattering through an interferometer
at constant speed $v_{F}$. The two beamsplitters $A$ and $B$ connect
the fermion fields $\hat{\psi}_{\alpha}$ of the input ($\alpha=1,2$)
and output ($\alpha=3,4$) channels to those of the left and right
arm ($\alpha=L,R$), which we take to be of equal length $l$:

\begin{eqnarray}
\hat{\psi}_{L}(0,t) & = & r_{A}\hat{\psi}_{1}(0,t)+t_{A}\hat{\psi}_{2}(0,t)\label{psiL}\\
\hat{\psi}_{R}(0,t) & = & t_{A}\hat{\psi}_{1}(0,t)+r_{A}\hat{\psi}_{2}(0,t)\label{psiR}\\
\hat{\psi}_{3}(l,t) & = & r_{B}e^{i\phi}\hat{\psi}_{L}(l,t)+t_{B}\hat{\psi}_{R}(l,t)\label{psi3}\\
\hat{\psi}_{4}(l,t) & = & t_{B}e^{i\phi}\hat{\psi}_{L}(l,t)+r_{B}\hat{\psi}_{R}(l,t)\label{psi4}\end{eqnarray}
The transmission (reflection) amplitudes $t_{A/B}\,(r_{A/B})$ fulfill
$t_{j}^{*}r_{j}=-t_{j}r_{j}^{*}$ due to unitarity, and we have included
the Aharonov-Bohm phase difference $\phi$. The input fields $\alpha=1,2$
obey $\left\langle \psi_{\alpha}^{\dagger}(0,0)\psi_{\alpha}(0,t)\right\rangle =\int_{-k_{c}}^{k_{c}}(dk)\, f_{\alpha k}e^{-iv_{F}kt}$
(note $\hbar=1$), with a band-cutoff $k_{c}$. We use the notation
$(dk)\equiv dk/(2\pi)$. 

The particles are assumed to have no intrinsic interaction, but are
subject to an external free bosonic quantum field $\hat{V}$ (linear
bath) during their passage through the arms $L,R$: $\hat{H}_{{\rm int}}=\sum_{\lambda=L,R}\int dx\,\hat{V}_{\lambda}(x)\hat{n}_{\lambda}(x)$
with $\hat{n}_{\lambda}(x)=\hat{\psi}_{\lambda}^{\dagger}(x)\hat{\psi}_{\lambda}(x)$. 

\emph{General expressions for current and shot noise} - We focus on
the current going into output port $3$, which is related to the density:
$\hat{I}(t)=ev_{F}\hat{n}_{3}(t)$ with $\hat{n}_{3}(t)=\hat{\psi}_{3t}^{\dagger}\hat{\psi}_{3t}$,
where we take fields $\hat{\psi}_{\alpha t}=\hat{\psi}_{\alpha}(l,t)$
at the position of the final beamsplitter B (except where noted otherwise).
In the following we set $e=v_{F}=1$, except where needed for clarity.
We have

\begin{equation}
\left\langle \hat{I}\right\rangle =R_{B}\left\langle \hat{\psi}_{L}^{\dagger}\hat{\psi}_{L}\right\rangle +T_{B}\left\langle \hat{\psi}_{R}^{\dagger}\hat{\psi}_{R}\right\rangle +e^{i\phi}t_{B}^{*}r_{B}\left\langle \hat{\psi}_{R}^{\dagger}\hat{\psi}_{L}\right\rangle +{\rm c.c.}.\label{CurrentAverage}\end{equation}
We have set $T_{B}=|t_{B}|^{2}$ and $R_{B}=1-T_{B}$. Without bath,
the interference term is given by $\left\langle \hat{\psi}_{R}^{\dagger}\hat{\psi}_{L}\right\rangle _{(0)}=r_{A}t_{A}^{*}\int(dk)\delta f_{k}=r_{A}t_{A}^{*}(eV/2\pi)$,
where we define $\delta f_{k}\equiv f_{1k}-f_{2k}$ and $\bar{f}_{k}\equiv(f_{1k}+f_{2k})/2$
for later use, and $eV=\mu_{1}-\mu_{2}$.

The zero-frequency current noise power is

\begin{equation}
S\equiv\int_{-\infty}^{+\infty}dt\,\left\langle \left\langle \hat{I}(t)\hat{I}(0)\right\rangle \right\rangle \,,\label{SDef}\end{equation}
where the double bracket denotes the irreducible part. The dependence
on $\phi$ and $T_{B},R_{B}$ is explicit, 

\begin{eqnarray}
 &  & S=R_{B}T_{B}C_{0}+R_{B}^{2}C_{0R}+T_{B}^{2}C_{0T}+\nonumber \\
 &  & 2\textrm{Re}\left[e^{i\phi}(t_{B}^{*}r_{B})(R_{B}C_{1R}+T_{B}C_{1T})-e^{2i\phi}T_{B}R_{B}C_{2}\right]\,.\label{S}\end{eqnarray}
with the coefficients following directly from inserting Eqs. (\ref{psi3})
into (\ref{SDef}), e.g. $C_{2}=\int dt\,\left\langle \left\langle \hat{\psi}_{Rt}^{\dagger}\hat{\psi}_{Lt}\hat{\psi}_{R0}^{\dagger}\hat{\psi}_{L0}\right\rangle \right\rangle $.
$C_{0(R/T)}$ are real-valued, the other coefficients may become complex.
The free values correspond to the result given by the well-known scattering
theory of shot noise of non-interacting fermions \cite{PartitionNoiseOriginal,BlanterReview}:

\begin{equation}
S_{(0)}=\int(dk)(f_{2k}+\delta f_{k}\mathcal{T})(1-(f_{2k}+\delta f_{k}\mathcal{T}))\,,\label{SNfree}\end{equation}
where $\mathcal{T}(\phi)=T_{A}T_{B}+R_{A}R_{B}+2t_{A}^{*}r_{A}t_{B}^{*}r_{B}\cos(\phi)$
is the transmission probability from $1$ to $3$. 

\emph{Symmetries of shot noise} - For our model, the full shot noise
power $S$ may be shown to be invariant under each of the following
transformations, if the bath couples equally to both arms of the interferometer:
(i) $t_{A}\leftrightarrow r_{A},\,\phi\mapsto-\phi$ (ii) $V\mapsto-V,\,\phi\mapsto-\phi$
(iii) $t_{B}\leftrightarrow r_{B}$. As a consequence, $C_{1T}=-C_{1R}$.
Note that the free result (\ref{SNfree}) is invariant under $\phi\mapsto-\phi$
and $V\mapsto-V$ separately, but these symmetries may be broken by
a bath-induced phase-shift.

\emph{Equations of motion} - We start from Heisenberg's equations
of motion for the fermions and the bath. The fermion field in each
arm obeys (omitting the index $L/R$ for now):

\begin{equation}
i(\partial_{t}-v_{F}\partial_{x})\hat{\psi}(x,t)=\int dx'K(x-x')\hat{V}(x',t)\hat{\psi}(x',t)\,,\label{PsiEqMotion}\end{equation}
where $\hat{V}$ evolves in presence of the interaction, see below.
We must consider states within a finite band, thus $K(x-x')=\{\hat{\psi}(x),\hat{\psi}^{\dagger}(x')\}\neq\delta(x-x')$.
Nevertheless, for the purpose of our subsequent leading-order approximation,
it turns out we can replace the right-hand side by $\hat{V}(x,t)\hat{\psi}(x,t)$
(neglecting, e.g., velocity-renormalization in higher orders). The
corresponding formal solution describes the accumulation of a random
phase:

\begin{eqnarray}
\hat{\psi}(x,t) & = & \hat{T}\exp\left[-i\int_{t_{0}}^{t}dt_{1}\,\hat{V}(x-v_{F}(t-t_{1}),t_{1})\right]\times\nonumber \\
 &  & \hat{\psi}(x-v_{F}(t-t_{0}),t_{0})\,.\label{SolvedPsiEq}\end{eqnarray}
In contrast to the case of classical noise \cite{unserPRL}, the field
$\hat{V}$ contains the response to the fermion density, in addition
to the homogeneous solution $\hat{V}_{(0)}$ of the equations of motion
(i.e. the free fluctuations): 

\begin{equation}
\hat{V}(x,t)=\hat{V}_{(0)}(x,t)+\int_{-\infty}^{t}dt'\, D^{R}(x,t,x',t')\hat{n}(x',t')\,.\label{SolvedVEq}\end{equation}
Here $D^{R}$ is the unperturbed retarded bath Green's function, $D^{R}(1,2)\equiv-i\theta(t_{1}-t_{2})\left\langle [\hat{V}(1),\hat{V}(2)]\right\rangle $,
where $\hat{V}$-correlators refer to the free field. This step is
analogous to the derivation of an operator quantum Langevin equation\cite{Weiss}
for any linear bath. 

Accounting for cross-correlations between the fluctuations in both
arms ({}``vertex-corrections'') is straightforward for a geometry
with symmetric coupling to parallel arms at a distance $d$ (assuming
$d\ll l$). Then, in the following results, $\left\langle \hat{V}\hat{V}\right\rangle =\left\langle \hat{V}_{L}\hat{V}_{L}\right\rangle -\left\langle \hat{V}_{L}\hat{V}_{R}\right\rangle $
and $D^{R}=D_{LL}^{R}-D_{LR}^{R}$. These correlators derive from
the threedimensional version, e.g. $\left\langle \hat{V}_{L}(x,t)\hat{V}_{R}(x',t)\right\rangle =\left\langle \hat{V}(x,y+d,z,t)\hat{V}(x',y,z,t')\right\rangle $.

\begin{figure}
\includegraphics[%
  width=1.0\columnwidth]{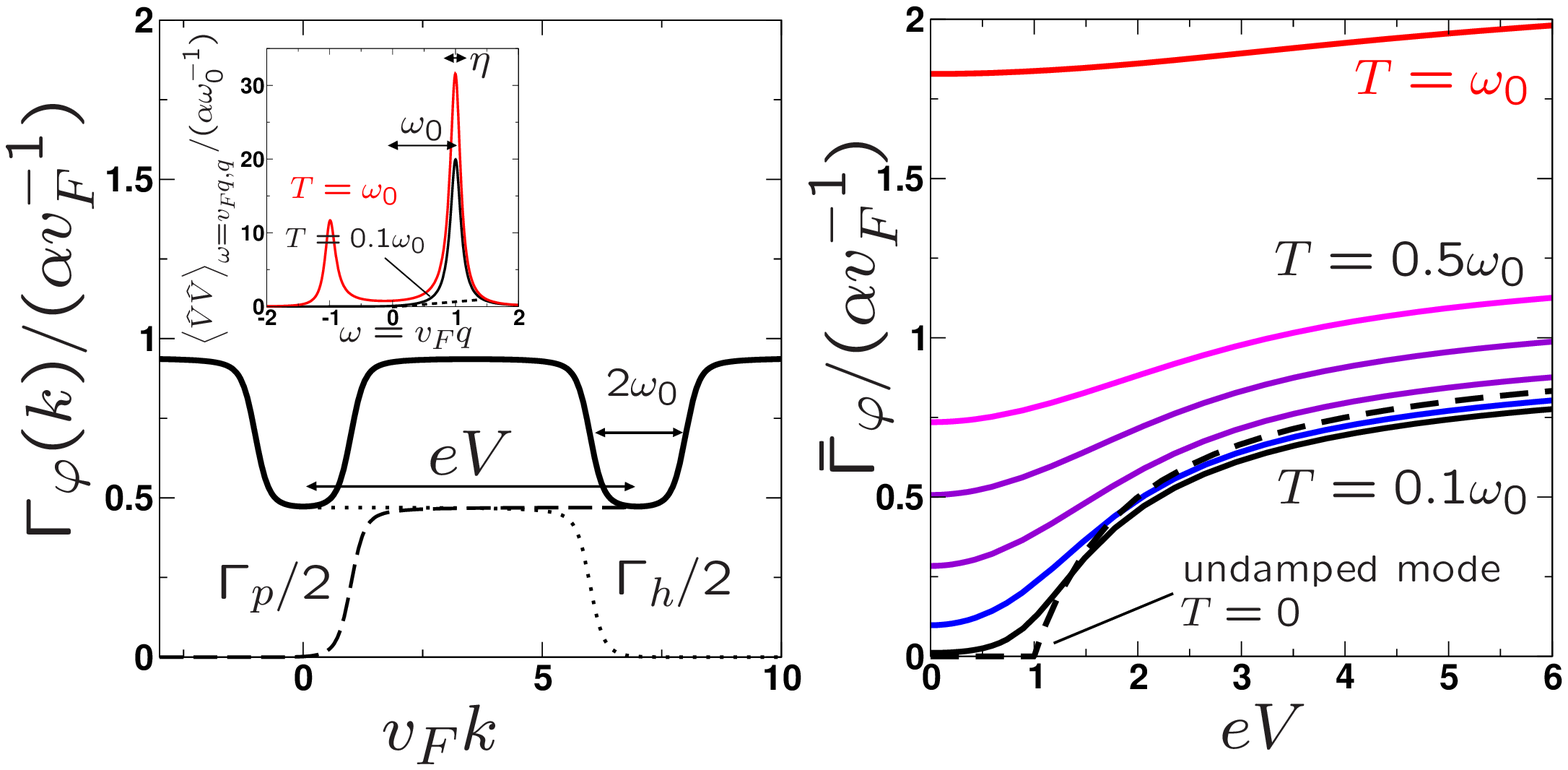}

\caption{\label{cap:Left:-Energy-resolved-dephasing}Left: Energy-resolved
dephasing rate for a sample bath spectrum (damped phonon mode, inset).
Right: Voltage-dependent energy-averaged dephasing rate for different
temperatures. Energies in units of $\omega_{0}$.}
\end{figure}

\emph{Interference term, renormalized phase shift and dephasing rate}
- In order to obtain the current or the noise, we expand the exponential
(\ref{SolvedPsiEq}) to second order, insert the formal solution (\ref{SolvedVEq}),
and perform Wick's averaging over fermion fields, while implementing
a {}``Golden Rule approximation'', i.e. keeping only terms linear
in the time-of-flight $\tau$. Then we obtain the following leading
correction to the interference term: 

\begin{equation}
\delta\left\langle \hat{\psi}_{R}^{\dagger}\hat{\psi}_{L}\right\rangle =r_{A}t_{A}^{*}\int(dk)\delta f_{k}[i\delta\bar{\varphi}(k)-\Gamma_{\varphi}(k)\tau]\label{InterferenceTerm}\end{equation}
Here the effective average $k$-dependent phase shift induced by coupling
to the bath is

\begin{equation}
\delta\bar{\varphi}(k)=\tau(R_{A}-T_{A})\int(dq)({\rm Re}D_{q,q}^{R}-D_{0,0}^{R})\delta f_{k-q}\,,\label{PhaseShift}\end{equation}
which vanishes for $T_{A}=1/2$, since then there is complete symmetry
between both arms. The interference term is suppressed according to
the total dephasing rate $\Gamma_{\varphi}(k)=\Gamma_{\varphi}^{L}(k)+\Gamma_{\varphi}^{R}(k)$,
with

\begin{eqnarray}
 &  & \Gamma_{\varphi}^{L}(k)=\int(dq)\,\left[\frac{1}{2}\left\langle \hat{V}\hat{V}\right\rangle _{q,q}+\textrm{Im}D_{q,q}^{R}f_{Lk-q}\right]\nonumber \\
 &  & =-\frac{1}{2}\int(dq)\,\textrm{Im}D_{q,q}^{R}\left[\coth\frac{\beta q}{2}-(f_{Lk-q}-f_{Lk+q})\right]\,.\label{Gammaphi}\end{eqnarray}
The {}``back-action'' $\propto D^{R}$ is crucial, since it introduces
the nonequilibrium Fermi functions ($f_{L}=R_{A}f_{1}+T_{A}f_{2}$,
$f_{R}=T_{A}f_{1}+R_{A}f_{2}$) which capture the physics of Pauli
blocking: Large energy transfers $v_{F}|q|\gg eV,T$ are forbidden
for states $k$ within the transport region. As a result, the interference
contrast becomes perfect for $V,T\rightarrow0$. The dephasing rate
is the sum of particle- and hole-scattering rates\cite{JvDAndMe},
$\Gamma_{\varphi}^{L}=(\Gamma_{p}^{L}+\Gamma_{h}^{L})/2$, with $\Gamma_{p}^{L}(k)=\int(dq)\,\left\langle \hat{V}\hat{V}\right\rangle _{q,q}(1-f_{Lk-q})$
and $\Gamma_{h}^{L}(k)=\int(dq)\,\left\langle \hat{V}\hat{V}\right\rangle _{q,q}f_{Lk+q}$.
This is because both processes destroy the superposition of many-particle
states that is created when a particle passes through the first beam
splitter, entering the left or the right arm (kets in Fig. \ref{cap:Schematic-of-the},
right). For linear transport, $f_{Lk-q}-f_{Lk+q}\rightarrow-\tanh(\beta(k-q)/2)$
under the integral, leading to the result well known in the theory
of weak localization\cite{WLdephasingDiags}, where ballistic motion
in our case ($\omega=v_{F}q$) is replaced by diffusion. Fig. \ref{cap:Left:-Energy-resolved-dephasing}
displays $\Gamma_{\varphi}(k)$ and $\bar{\Gamma}_{\varphi}=(eV)^{-1}\int dk\,\delta f_{k}\Gamma_{\varphi}(k)$
for the illustrative example of a damped optical phonon mode, $D_{q,\omega}^{R}=\alpha[(\omega-\omega_{0}+i\eta)^{-1}-(\omega+\omega_{0}+i\eta)^{-1}]$. 

\emph{Shot noise correction} - In the same manner, after a straightforward
but lengthy calculation, we arrive at the following leading-order
corrections to the noise power $S$ (again keeping only terms $\propto\tau^{1}$):

\begin{eqnarray}
 &  & \frac{\delta C_{0}}{4\tau R_{A}T_{A}}=-\int(dk)(dq)\,{\rm Im}D_{q,q}^{R}\times\nonumber \\
 &  & \left[\delta f_{k}\delta f_{k+q}(\bar{f}_{k+q}-\bar{f}_{k})+\right.\nonumber \\
 &  & \left.(f_{1k}^{2}+f_{2k}^{2})\bar{f}_{k+q}-(f_{1k+q}^{2}+f_{2k+q}^{2})\bar{f}_{k}\right]+\nonumber \\
 &  & \int(dk)(dq)\left\langle \hat{V}\hat{V}\right\rangle _{q,q}\times\nonumber \\
 &  & [(f_{1k+q}-f_{1k})(1-f_{1k})+(f_{2k+q}-f_{2k})(1-f_{2k})]\nonumber \\
 &  & +(eV/2\pi)^{2}\left\langle \hat{V}\hat{V}\right\rangle _{0,0}\,,\label{dC0}\end{eqnarray}
while $\delta C_{0R/T}=0$. We have $\delta C_{1T}=-\delta C_{1R}$
with

\begin{eqnarray}
 &  & {\rm Re}\frac{\delta C_{1R}}{\tau r_{A}t_{A}^{*}(R_{A}-T_{A})}=\int(dk)(dq){\rm Im}D_{q,q}^{R}\times\nonumber \\
 &  & [\delta f_{k}\delta f_{k+q}(\bar{f}_{k+q}+3\bar{f}_{k}-2)+\delta f_{k+q}^{2}\bar{f}_{k}-\delta f_{k}^{2}\bar{f}_{k+q}]+\nonumber \\
 &  & \left[\int(dq)\left\langle \hat{V}\hat{V}\right\rangle _{q,q}\right]\left[\int(dk)\delta f_{k}^{2}\right]\label{dC1Rreal}\end{eqnarray}
and

\begin{eqnarray}
 &  & {\rm Im}\frac{\delta C_{1R}}{\tau r_{A}t_{A}^{*}}=\int(dk)(dq){\rm Re}D_{q,q}^{R}\times\nonumber \\
 &  & \left[-\delta f_{k}\delta f_{k+q}(\delta f_{k+q}+2\delta f_{k})(T_{A}-R_{A})^{2}/2+\right.\nonumber \\
 &  & 2R_{A}T_{A}\delta f_{k}\delta f_{k+q}^{2}+\delta f_{k+q}\bar{f}_{k}(3-2\bar{f}_{k})\nonumber \\
 &  & \left.-\delta f_{k}\bar{f}_{k+q}+2\bar{f}_{k}\bar{f}_{k+q}(\delta f_{k}-\delta f_{k+q})\right]\nonumber \\
 &  & +D_{0,0}^{R}(eV/2\pi)\int(dk)\times\nonumber \\
 &  & \left[\delta f_{k}^{2}(\frac{3}{2}(T_{A}^{2}+R_{A}^{2})-5R_{A}T_{A})-2\bar{f}_{k}(1-\bar{f}_{k})\right]\,.\label{dC1Rimag}\end{eqnarray}
Finally

\begin{eqnarray}
 &  & \frac{{\rm Re}\delta C_{2}}{2\tau R_{A}T_{A}}=(eV/2\pi)^{2}\left\langle \hat{V}\hat{V}\right\rangle _{0,0}-\nonumber \\
 &  & 2\int(dk)(dq)\,{\rm Im}D_{q,q}^{R}\bar{f}_{k}\delta f_{k+q}(\delta f_{k}+\delta f_{k+q})-\nonumber \\
 &  & \int(dk)(dq)\left\langle \hat{V}\hat{V}\right\rangle _{q,q}\delta f_{k}(\delta f_{k}+\delta f_{k+q})\label{dC2real}\end{eqnarray}
and

\begin{eqnarray}
 &  & {\rm Im}\delta C_{2}=-4R_{A}T_{A}\tau\int(dk)\delta f_{k}^{2}\delta\bar{\varphi}_{k}\,.\label{dC2imag}\end{eqnarray}
\emph{Discussion of shot noise} - We note that $\delta S\rightarrow0$
for $V\rightarrow0$ at arbitrary $T$, i.e. there is no Nyquist noise
correction. This is plausible, as $S_{(0)}(V=0)$ does not depend
on $\phi$ and thus is not affected by phase fluctuations. In the
case of purely classical noise (where $\hat{V}\mapsto V_{{\rm cl}},\, D^{R}\mapsto0$),
we found a finite $\phi$-independent Nyquist correction\cite{unserPRL},
but this is due to heating by a bath which is nominally at infinite
temperature (according to the fluctuation-dissipation theorem FDT,
applied to the case $D^{R}=0$).

\begin{figure}
\includegraphics[%
  width=1.0\columnwidth]{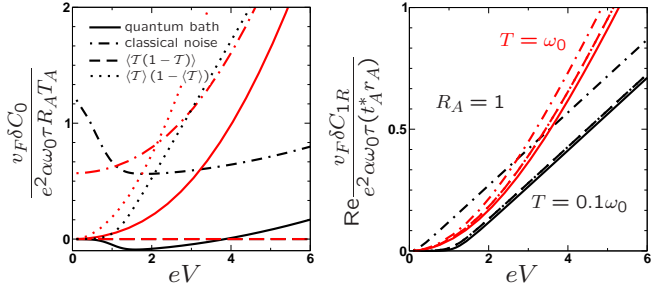}

\caption{\label{cap:Shot-noise-corrections}Shot noise corrections $\delta C_{0}$
and $\delta C_{1}$ as function of voltage and temperature (for the
bath spectrum in Fig. \ref{cap:Left:-Energy-resolved-dephasing}).
Comparison with classical noise, $\left\langle \mathcal{T}(1-\mathcal{T})\right\rangle $
and $\left\langle \mathcal{T}\right\rangle (1-\left\langle \mathcal{T}\right\rangle )$,
see text.}
\end{figure}

As expected, the $\phi$-dependence in the shot noise is suppressed
(\ref{S}): $|{\rm Re}C_{2}|$ and $|{\rm Re}C_{1R}/(r_{A}t_{A}^{*})|$
decrease.

At large $V$ (larger than the bath spectrum cutoff), there is a contribution
rising like $V^{2}$ due to time-dependent conductance fluctuations
($\left\langle \hat{V}\hat{V}\right\rangle _{0,0}$ terms in Eqs.
(\ref{dC0}) and (\ref{dC2real})), corresponding to the leading order
of {}``$S_{{\rm cl}}$'' in Refs. \cite{unserPRL}. The average
phase shift grows like $V$ (increasing density-difference between
the arms), producing a $V^{2}$-behaviour in the phase-shift terms
(\ref{dC1Rimag}) and (\ref{dC2imag}) at large voltages, whereas
Eq. (\ref{dC1Rreal}) rises like $V^{1}$ (Fig. \ref{cap:Shot-noise-corrections},
right). 

There are two peculiar features of the shot noise correction. First,
the coefficient ${\rm Im}\delta C_{2}$ of Eq. (\ref{dC2imag}) is
twice as large as expected from Eq. (\ref{S}) on the grounds of a
simple phase shift. Second, Eq. (\ref{dC1Rimag}) does not vanish
for $T_{A}=1/2$, such that there remains a $\phi\leftrightarrow-\phi$-asymmetry
in $\delta S$ even when both arms are completely symmetric and there
is no phase shift in the current pattern ($\delta\bar{\varphi}_{k}=0$).
Only the additional constraint $T_{B}=1/2$ will guarantee the $\phi$-symmetry. 

Both features arise because the phase shift fluctuates, due to the
density fluctuations in both arms. Restricting attention to the $k$-independent
part for ease of the discussion, we may interpret the phase shift
as an operator depending on the densities, schematically $\delta\hat{\varphi}[\hat{n}_{L/R}]$.
It is correlated with the output current, $\left\langle (\delta\hat{\varphi}(t)-\delta\bar{\varphi})(\hat{I}(0)-\bar{I})\right\rangle \neq0$,
leading to an extra shot noise contribution and (together with the
$k$-dependent part) accounting for the extra factor of two in Eq.
(\ref{dC2imag}), as well as the fact that $T_{A}=1/2$ is not enough
to obtain a $\phi$-symmetric shot noise (since the correlator $\left\langle \delta\hat{\varphi}\hat{I}\right\rangle $
depends on $T_{B}$ as well).

\emph{Comparison with simpler approaches} - The limit of classical
noise (treated to all orders in Refs. \cite{unserPRL}) is recovered
by setting $D^{R}=0$ in Eqs. (\ref{dC1Rreal})-(\ref{dC2imag}) and
inserting the symmetrized correlator $\left\langle V_{{\rm cl}}V_{{\rm cl}}\right\rangle =\left\langle \left\{ \hat{V},\hat{V}\right\} \right\rangle /2$.
For $\delta C_{0}$, Eq. (\ref{dC0}) has to be replaced by \begin{eqnarray}
 &  & \frac{\delta C_{0}^{{\rm cl}}}{\tau}=2\int(dk)(dq)\left\langle V_{{\rm cl}}V_{{\rm cl}}\right\rangle _{q,q}\times\nonumber \\
 &  & [(f_{Lk+q}-f_{Lk})(1-f_{Rk})+(f_{Rk+q}-f_{Rk})(1-f_{Lk})]+\nonumber \\
 &  & 4R_{A}T_{A}\left[\int(dk)\,\delta f_{k}\right]^{2}\left\langle V_{{\rm cl}}V_{{\rm cl}}\right\rangle _{0,0}\,,\label{dC0classical}\end{eqnarray}
since (\ref{dC0}) was already simplified using the FDT. Eq. (\ref{dC0classical})
contains the finite Nyquist noise correction discussed above (Fig.
\ref{cap:Shot-noise-corrections}). It is impossible to recover the
full quantum noise result by inserting some suitably modified classical
noise correlator $\left\langle V_{{\rm cl}}V_{{\rm cl}}\right\rangle $
(containing Fermi functions for Pauli blocking, effectively like Refs.
\cite{WLdephasingClassical}), since this cannot yield the important
phase shift terms, although it works for the dephasing rate. The conductance
fluctuations are correctly captured even by the classical approach.

One can also implement a phenomenological ansatz for the partition
noise, of the form $\left\langle \mathcal{T}(\phi+\varphi)(1-\mathcal{T}(\phi+\varphi))\right\rangle _{\varphi}$
or $\left\langle \mathcal{T}(\phi+\varphi)\right\rangle _{\varphi}\left\langle 1-\mathcal{T}(\phi+\varphi)\right\rangle _{\varphi}$
(as used for the interpretation in Ref. \cite{HeiblumEtAl}, see Refs.
\cite{unserPRL}). More precisely, we introduce fluctuations $\phi\mapsto\phi+\delta\varphi_{k}$
into the scattering theory result, Eq. (\ref{SNfree}), and average
according to either $\left\langle \mathcal{TT}\right\rangle $ or
$\left\langle \mathcal{T}\right\rangle \left\langle \mathcal{T}\right\rangle $,
assuming Gaussian variables $\delta\varphi_{k}$ , with $\delta\bar{\varphi}_{k}$
taken from Eq. (\ref{PhaseShift}) and $\left\langle \delta\varphi_{k}^{2}\right\rangle =2\tau\Gamma_{\varphi}(k)$.
This procedure is designed to reproduce the correct average current.
The resulting phase shift terms are the same for both variants, although
they deviate from the quantum solution due to the influence of phase
shift fluctuations (see above). The absence of a Nyquist correction
is trivially reproduced, while the conductance fluctuations are missed
(Fig. \ref{cap:Shot-noise-corrections}, left).

Regarding future experiments following Ref. \cite{HeiblumEtAl}, we
propose to check the following points, which are independent of the
assumed microscopic quantum noise correlator: (i) the shot noise symmetries
(ii) the dependence of $S$ on $T_{B}$ and $\phi$ (only $e^{i\phi}$,
$e^{2i\phi}$ contributions present) predicted in Eq. (\ref{S}) (iii)
the general $T_{A}$-dependence of the shot noise corrections at good
visibility (comparing to Eqs. (\ref{dC0})-(\ref{dC2imag})) (iv)
Observing different phase-shifts of the $e^{i\phi}$ and $e^{2i\phi}$-contributions
to $S$ would be particularly interesting, as this feature is not
present in any simple phenomenological (or classical noise) model.

\emph{Conclusions} - We have analyzed the influence of a quantum bath
on the current and the shot noise in a fermionic Mach-Zehnder interferometer,
employing equations of motion and perturbation theory. We have found
a dephasing rate that is a nonequilibrium generalization of the well-known
weak-localization result, with scattering restricted by nonthermal
distribution functions. The crucial Pauli blocking effects are described
as a consequence of the backaction of the bath onto the system. Our
result for the current noise cannot be reproduced by inserting any
effective classical noise correlator or by any of various simpler
models. We have found and explained interesting extra shot noise contributions
which are due to the phase shift depending on the fluctuating distribution
functions within the arms, that are correlated with the output current
itself. The present approach lends itself naturally to a systematic
extension to higher orders, as well as possible variations such as
the inclusion of curvature within the interferometer arms, calculations
beyond the Golden Rule approximation, and the analysis of current
cross-correlators. 

\emph{Acknowledgments} - I thank S. M. Girvin, A. A. Clerk, C. Bruder,
J. v. Delft, T. Novotn\'{y} and V. Fal'ko for illuminating discussions,
and the DFG for financial support (MA 2611/1-1).

\end{document}